\documentclass[aps,prb,twocolumn,superscriptaddress,10pt,showkeys,showpacs]{revtex4-1}
\usepackage{graphicx}
\usepackage{color}
\usepackage{url}
\usepackage[normalem]{ulem}

\begin{document}

\title{Interpretation of monoclinic hafnia valence electron energy loss spectra by TDDFT}
%\title{Nanospectroscopic properties of hafnium oxides: HRTEM-VEELS measurements and TDDFT simulations}
\author{L. Hung}
\altaffiliation[Current address: ]{NIST Center for Neutron Research, National Institute of Standards and Technology, Gaithersburg, Maryland 20899-6102, USA}
\affiliation{Laboratoire des Solides Irradi\'es, \'Ecole Polytechnique, CNRS, 
CEA, Universit\'e Paris-Saclay, F-91128 Palaiseau, France}
\affiliation{European Theoretical Spectroscopy Facility (ETSF)}
\author{C. Guedj}
\affiliation{Univ. Grenoble Alpes, F-38000 Grenoble, France}
\affiliation{CEA, LETI, MINATEC campus, F38054 Grenoble, France}
\author{N. Bernier}
\affiliation{Univ. Grenoble Alpes, F-38000 Grenoble, France}
\affiliation{CEA, LETI, MINATEC campus, F38054 Grenoble, France}
\author{P. Blaise}
\affiliation{Univ. Grenoble Alpes, F-38000 Grenoble, France}
\affiliation{CEA, LETI, MINATEC campus, F38054 Grenoble, France}
\affiliation{European Theoretical Spectroscopy Facility (ETSF)}
\author{V. Olevano}
\affiliation{Univ. Grenoble Alpes, F-38000 Grenoble, France}
\affiliation{CNRS, Institut N\'eel, F-38042 Grenoble, France}
\affiliation{European Theoretical Spectroscopy Facility (ETSF)}
\author{F. Sottile}
\affiliation{Laboratoire des Solides Irradi\'es, \'Ecole Polytechnique, CNRS, 
CEA, Universit\'e Paris-Saclay, F-91128 Palaiseau, France}
\affiliation{European Theoretical Spectroscopy Facility (ETSF)}

\date{\today}

\begin{abstract}
We present the valence electron energy-loss spectrum and the dielectric function of monoclinic hafnia (m-HfO$_2$) obtained from time-dependent density-functional theory (TDDFT) predictions and compared to energy-filtered spectroscopic imaging measurements in a high-resolution transmission-electron microscope. 
Fermi's Golden Rule density-functional theory (DFT) calculations can capture the qualitative features of the energy-loss spectrum, but we find that TDDFT, which accounts for local-field effects, provides nearly quantitative agreement with experiment.
Using the DFT density of states and TDDFT dielectric functions, we characterize the excitations that result in the m-HfO$_2$ energy loss spectrum.
The sole plasmon occurs between 13-16 eV, although the peaks $\sim$28 and above 40 eV are also due to collective excitations.
We furthermore elaborate on the first-principles techniques used, their accuracy, and remaining discrepancies among spectra. 
More specifically, we assess the influence of Hf semicore electrons (5$p$ and 4$f$) on the energy-loss spectrum, and find that the inclusion of transitions from the 4$f$ band damps the energy-loss intensity in the region above 13 eV.
We study the impact of many-body effects in a DFT framework using the adiabatic local-density approximation (ALDA) exchange-correlation kernel, as well as from a many-body perspective using a $GW$-derived electronic structure to account for self-energy corrections.
These results demonstrate some cancellation of errors between self-energy and excitonic effects, even for excitations from the Hf $4f$ shell.
We also simulate the dispersion with increasing momentum transfer for plasmon and collective excitation peaks.
\end{abstract}

\pacs{77.22.Ch, 79.20.Uv, 71.15.Mb, 71.15.Qe}
\keywords{HfO$_2$, hafnia, monoclinic, TEM, VEELS, EELS, DFT, TDDFT}

\maketitle

\section{Introduction}

Hafnia-based dielectric materials are among the most promising and extensively studied high-$\kappa$ materials, due to HfO$_2$'s high permittivity, relatively wide band gap, and compatibility with Si that make it useful for applications in micro- and nano-electronics.\cite{ChoiReview,RobertsonReview}
To obtain improved dielectric and stability properties, much research focuses on doping HfO$_2$ or studying alternative interfaces and phases.
However, an accurate characterization of the parent HfO$_2$ material can benefit both fundamental knowledge and technological advances.

To better understand the dielectric properties of HfO$_2$,
we study $P2_1/c$ monoclinic HfO$_2$ (m-HfO$_2$) by experiment and first-principles theory.
We acquire energy-loss spectra with both good energy resolution and nanometer-scale spatial resolution using valence electron energy-loss spectroscopy (VEELS) combined with high-resolution transmission electron microscopy (HRTEM).\cite{Abajo,VerbeeckVanDyck}
To simulate VEELS, we use time-dependent density-functional theory (TDDFT)\cite{RungeGross,GrossKohn,ZangwillSoven}.
A complete theoretical description of all dielectric properties must take into account physical processes involved in both single-particle and collective (e.g., plasmon) excitations, and include electron-hole (excitonic) and electron-electron (self-energy) interaction effects.
For the energy range of VEELS, the prominent features of spectra are typically caused by collective excitations.
For such excitations, TDDFT has demonstrated good agreement with experiment, due to the compensation of self-energy and excitonic effects.\cite{OlevanoReining01}
Notably, the TDDFT and experimental energy-loss spectra of ZrO$_2$, which is isomorphous to HfO$_2$, are in quantitative agreement.\cite{Dash}
In this work, we compare TDDFT-predicted energy-loss spectra for m-HfO$_2$ with HRTEM-VEELS measurements for the dual purposes of
(1) characterizing the peaks in the energy-loss spectrum and
(2) understanding discrepancies between theory and experiment.
In addition to addressing local-field (LF) effects, we examine the contributions from localized semicore wave functions (especially the 4$f$ electrons of Hf), many-body exchange-correlation effects at the adiabatic local-density approximation (ALDA) and at the $GW$ level, and nonzero momentum transfer.
We discuss some results in the context of anisotropic effects, which were already presented in previous work,\cite{APL2014} but most of our results here refer to averaged spectra corresponding to polycrystalline m-HfO$_2$.

This paper is organized as follows: Sec.~\ref{sec:methods} describes our theoretical and experimental methods.
In Sec.~\ref{sec:interpretation}, we characterize the excitations reflected in the features of experimental and theoretical energy-loss spectra (and the corresponding dielectric functions).
In Sec.~\ref{sec:details}, we report the theoretical spectra obtained via first-principles TDDFT calculations, separating out the various contributions mentioned above (LF, semicore electrons, exchange-correlation effects, and momentum transfer).
%To study many-body interactions, we consider both the influence of the local-density approximation (LDA) exchange-correlation functional, as well as electron-electron self-energy corrections to the electronic structure.
We summarize our results in Sec.~\ref{sec:conclusion}.

\begin{table}
	\caption{Unit cell lattice parameters for m-HfO$_2$ calculated by DFT-LDA and measured by HRTEM.}
	\label{lattice_parameters}
	\begin{ruledtabular}
	\begin{tabular}[c]{lllll}
		& $a$ [\AA] & $b$ [\AA] & $c$ [\AA] & $\beta$ [degrees]\\\hline
		DFT-LDA & 5.05 & 5.14 & 5.22 & 99.56 \\
		HRTEM   & 5.1156 & 5.1722 & 5.2948 & 99.259
	\end{tabular}
	\end{ruledtabular}
\end{table}

\section{Methods}
\label{sec:methods}
\subsection{Theory}

\newcommand{\rv}{\mathbf{r}}
\newcommand{\gv}{\mathbf{G}}
\newcommand{\qv}{\mathbf{q}}
\newcommand{\kv}{\mathbf{k}}

First-principles calculations are carried out within the framework of Kohn-Sham density-functional theory (DFT)\cite{HohenbergKohn,KohnSham} and TDDFT\cite{RungeGross,GrossKohn,ZangwillSoven} using a plane wave pseudopotential implementation.
For the scalar-relativistic Hf pseudopotential, the semicore 5$s$, 5$p$, and 4$f$ electrons are treated as valence (in addition to the 5$d$ and 6$s$ electrons) since they have relatively low binding energies ($\sim$64 eV, $\sim$30 eV, and $\sim$10 eV, respectively, in HfO$_2$) such that the 5$p$ and 4$f$ electrons contribute to the energy range of interest.
% We have chosen a core radius of 2 bohrs %(Francesco, check the input FHI98CP code).
A scalar-relativistic Hf pseudopotential with only 5$d$ and 6$s$ electrons as valence is also tested to demonstrate the influence of the semicore electrons.
All pseudopotentials are constructed using the Trouiller-Martins scheme.\cite{TroullierMartins}

The ground-state geometry and electronic structure are computed with DFT and the local-density approximation (LDA) exchange-correlation functional.
We use the ABINIT code\cite{abinit} with a kinetic energy cutoff of 150 Ha and a $4 \times 4 \times 4$ Monkhorst Pack k-point grid.
The lattice vectors and atomic positions of m-HfO$_2$ are optimized in DFT, and the resulting unit cell dimensions are in good agreement with our HRTEM measurements (Table~\ref{lattice_parameters}) and literature values.\cite{WangLi}

Using the DP code,\cite{dp} we compute the dielectric function and energy-loss spectra of m-HfO$_2$ via linear-response TDDFT as follows.
First, we represent the independent particle polarizability as a matrix in terms of reciprocal lattice vectors $\gv$ and $\gv'$, momentum transfer $\qv$, and energy $\omega$:
\[
  \chi^0_{\gv\gv'}(\qv,\omega) = 
    \sum_{n, m, \kv} (f_{m\kv} - f_{n\kv})
  \frac{
\tilde{\rho}_{nm\kv}(\qv,\gv)\tilde{\rho}^*_{nm\kv}(\qv,\gv')}
{\omega- (\epsilon_{n\kv}-\epsilon_{m\kv}) + i\eta},
\]
where $\eta$ an infinitesimal positive value, $\epsilon_{n\mathbf{k}}$ are the Bloch DFT Kohn-Sham eigenvalues and $f_{n\kv}$ their occupation,
$\tilde{\rho}_{nm\kv}(\qv,\gv) = \int d\rv \, \psi_{n\kv}(\rv)e^{-i(\qv+\gv)\cdot\rv}\psi^*_{m\kv-\qv}(\rv) $ 
is constructed from the DFT Kohn-Sham wave functions $\psi_{n\kv}(\rv)$,
with wave vectors $\mathbf{k}$ and momentum transfer $\qv$ lying within the first Brillouin zone, spatial coordinate $\rv$, and band indices $n$ and $m$.  The TDDFT full polarizability is then determined via the Dyson equation,
\begin{equation}
\chi = \chi^0 + \chi^0 \left( v+ f_{xc}\right) \chi,
\end{equation}
where $v$ is the Coulomb potential and $f_{xc}$ is the exchange-correlation kernel.  The full polarizability is related to the inverse dielectric function by $\varepsilon^{-1} = 1+ v\chi$, and the electron energy-loss function is $-\Im(\varepsilon^{-1})$.

The real and imaginary parts of the dielectric function ($\Re(\varepsilon)$ and $\Im(\varepsilon)$, respectively) are used to characterize features in the energy-loss spectra.
Features located at energy losses where the $\Re(\varepsilon)$ passes through zero (going from negative to positive) are collective excitations known as plasmons.
On the other hand, if the $\Im(\varepsilon)$ is large at the energy of some loss-spectra feature, the feature is attributed to single-particle inter-band excitations.
Finally, features occurring at energies where the $\Re(\varepsilon)$ is small but nonzero correspond to (non-plasmon) collective excitations.

Our calculations of dielectric functions and energy-loss spectra are converged after including $n=300$ bands, a basis of 9475 plane waves for the wave functions, and 329 plane waves for the dielectric matrices ($\chi^0$, $\chi$ and $\varepsilon$).
Here, we sample the Brillouin zone with $4 \times 4 \times 4$ k-point grids shifted in low-symmetry directions.\cite{Benedict}
Unless otherwise indicated, the spectra have a Gaussian broadening of $1.5$ eV to smooth the sampling error over the Brillouin zone and to take into account our experimental energy resolution of $1\sim1.3$ eV.
Also unless stated otherwise, the computed spectra correspond to vanishing momentum transfer ($q \rightarrow 0$), and are averaged over three reciprocal lattice directions, for comparison to measurements taken on polycrystalline samples.

By using a TDDFT framework, our simulations allow the inclusion of LF effects, which arise from anisotropies\cite{anisotropy} and local inhomogeneities in the material and are crucial in the description of the HfO$_2$ loss function.
To aid in interpreting the spectra, we also present results that leave out LF effects (NLF), by only using the diagonals of the $\chi^0$, $\chi$ and $\varepsilon$ matrices.
Such NLF calculations correspond to Fermi's Golden Rule DFT predictions.
Most of our calculations use TDDFT in the random-phase approximation (RPA), where $f_{xc}=0$.
In addition, we consider TDDFT using the ALDA exchange-correlation kernel (TDLDA).
We also evaluate the effect of the many-body electron-electron self-energy by applying a ``scissor operator" (SO) to the DFT electronic structure.
In contrast to starting from the Kohn-Sham DFT electronic structure (as is done in most our calculations), the SO shifts eigenvalues to roughly match $GW$ quasiparticle energies\cite{Jiang} while keeping wave functions unchanged.

\subsection{Experiment}

\begin{figure}
	\includegraphics[width=\columnwidth]{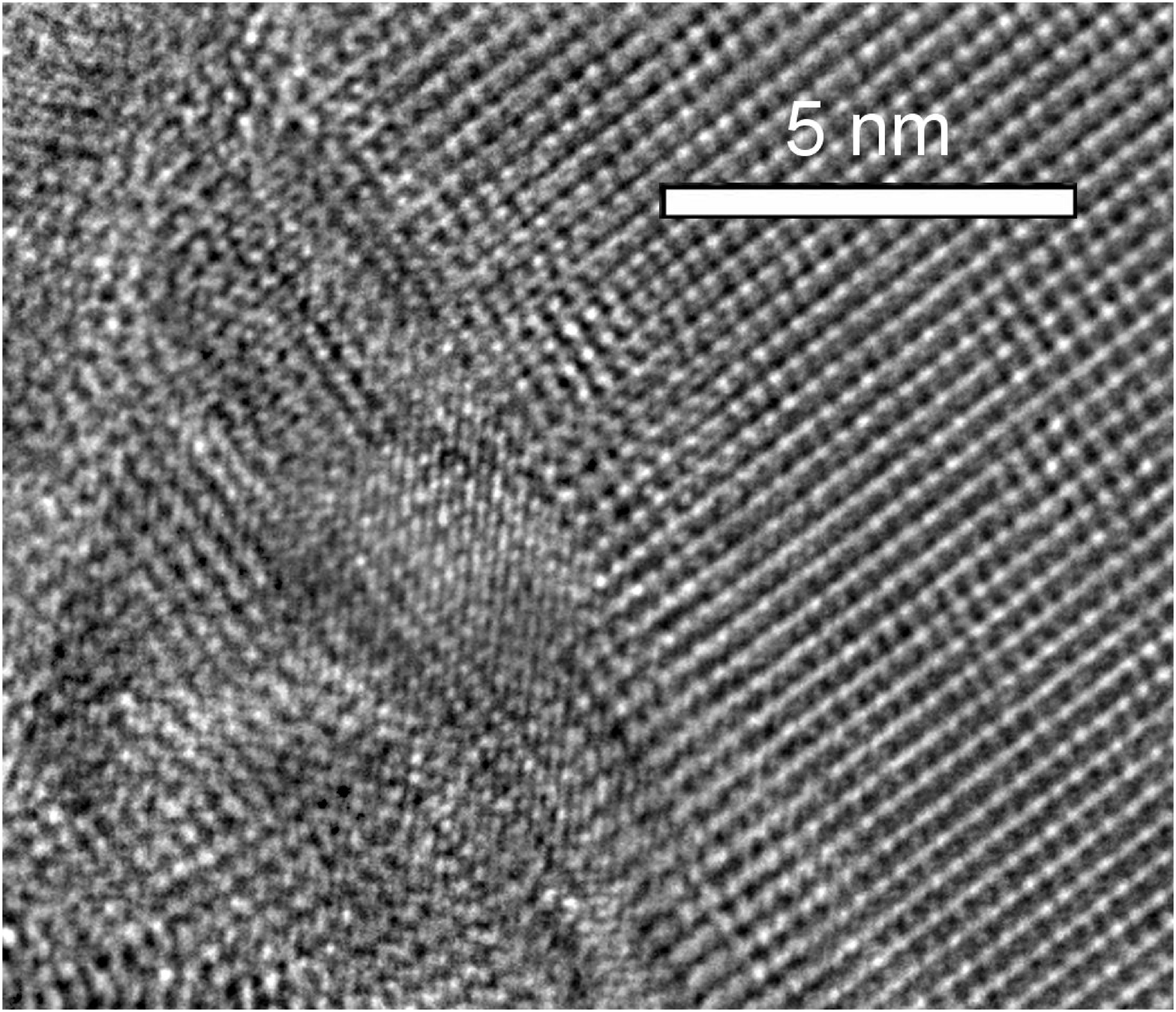}
	\caption{Cross-sectional HRTEM image of m-HfO$_2$ showing a polycrystalline area (left) and a single crystalline domain (right).
	}
	\label{tem}
\end{figure}

Our samples (previously described in Ref.~[\onlinecite{APL2014}]) consist of decananometric hafnia layers grown on 200 mm p-Si(100) wafers by atomic layer deposition (ALD) in a cleanroom environment dedicated to the semiconductor industry.
ALD samples are prepared with a Strata\texttrademark\ 400 DualBeam\texttrademark\ FIB/STEM system using Ga$^+$ ions energies ranging from 30 keV down to 2 keV.
An improvement in the quality of HRTEM-VEELS data is obtained by selective lift-off of superficial amorphous species by HF etching.
Transmission electron microscope (TEM) lamella thickness is optimized to avoid the need for multiple scattering deconvolution processing ($< 40$ nm) and to avoid excessive surface effects ($> 15$ nm).
The m-HfO$_2$ samples are polycrystalline with grains of varying size (Fig.~\ref{tem}), and appear to be stable under e-beam irradiation in the time scale of the measurements.
High grade m-HfO$_2$ commercial powders are also used for verification purposes.  

Cross-sectional electron nanospectroscopic imaging experiments are performed in a JEOL 2010 FEF TEM operated at 200 kV in high resolution mode.
The acquisition is performed in the energy filtered mode (EFTEM),\cite{EFTEM} by recording images from a selected energy-loss range from an omega filter with an energy step of 0.1 eV between each image acquisition.
In order to minimize experimental momentum dispersion, a nearly parallel configuration (nbed mode) is used, and the convergence angle is less than 0.2 mrad.
The lowest achievable collection angles is used, and the measured energy resolution is close to 1 eV.
Experimental data are corrected using the guidelines provided by Schaffer et al.\cite{SCHAFFER}
For verification, complementary results are obtained with the Cs-corrected FEI Titan microscope operated at 200 keV in STEM spectrum imaging \cite{STEMSI} and TEM modes.

VEELS spectra are measured for each pixel in a HRTEM image and selectively averaged to provide a spectrum representative of a random polycrystal.
The zero-loss (elastic) contribution is removed from a reference VEELS spectrum acquired simultaneously in the vacuum region closest to the measured region of interest.
The Kramers-Kronig analysis \cite{Egerton} is then performed on the single scattering distributions using classical routines available in the Digital Micrograph\texttrademark\ environment to provide complex permittivities, energy-loss functions and surface-loss functions.
Quantitative spectra are difficult to obtain because of the numerous sources of variability due to instrumentation, sample preparation and data analysis,
and about 100 million spectra are acquired over 20 different samples to obtain sufficient statistics.

\section{Interpreting energy-loss spectra}
\label{sec:interpretation}

\subsection{TDDFT vs. HRTEM-VEELS}
\begin{figure}
	\includegraphics[width=\columnwidth]{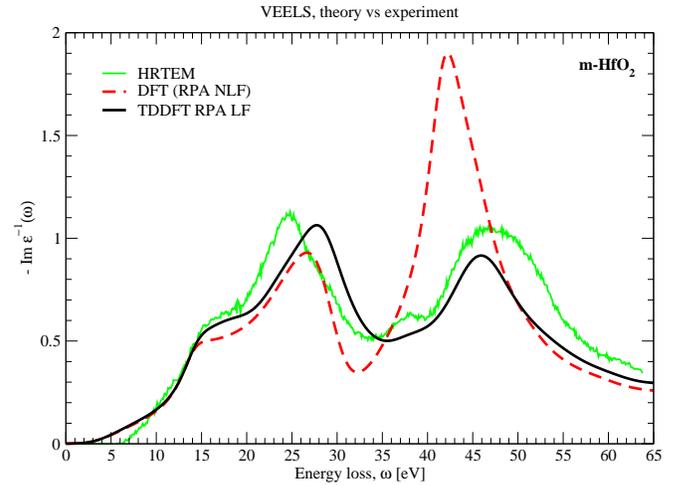}
	\caption{VEELS spectra for m-HfO$_2$. 
		Thin solid green line: HRTEM-VEELS averaged on a polycrystalline sample;
		dashed red line: DFT without LF effects;
		and solid black line: TDDFT RPA with LF effects.
        Theoretical curves have been convoluted with a Gaussian broadening of 1.5 eV.
	}
	\label{elf}
\end{figure}

\begin{figure}
	\includegraphics[width=\columnwidth]{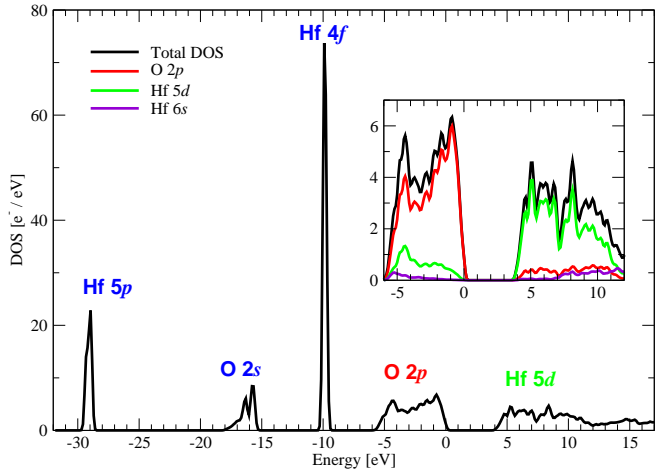}
	\caption{The density of states (DOS) for m-HfO$_2$.
		The inset shows the projected DOS near the band gap in greater detail.}
	\label{dos}
\end{figure}

In Fig.~\ref{elf} we show the HRTEM-VEELS spectrum and TDDFT RPA energy-loss spectra with and without LF effects.
Since the RPA NLF energy-loss spectrum is already in qualitative agreement with HRTEM-VEELS,
we use predictions at this level of theory to begin characterization of the m-HfO$_2$ energy-loss spectrum.
It is straightforward to determine which transitions contribute to each excitation using the DFT-LDA density of states (DOS) (Fig.~\ref{dos}).
The nature of the excitations (single-particle, plasmon, or collective) are characterized using the RPA NLF dielectric function.

\begin{figure}
	\includegraphics[width=\columnwidth]{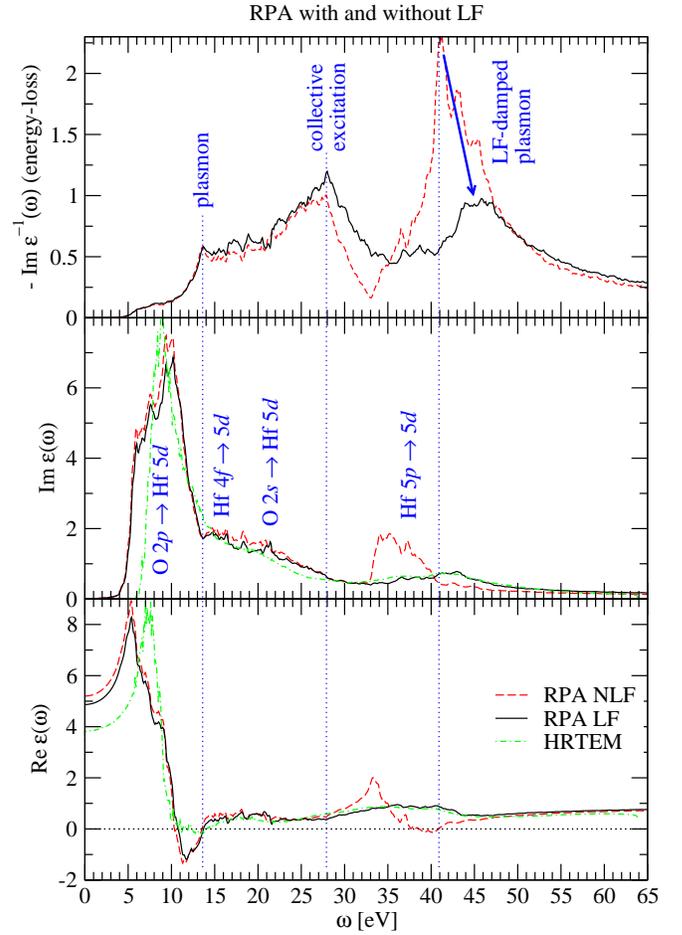}
	\caption{Energy-loss function (top), imaginary part of the dielectric function (middle), and the real part of the dielectric function (bottom) for m-HfO$_2$.
		Dashed red line: RPA without LF effects;
		solid black line: RPA with LF effects;
		and dot-dashed green line: HRTEM-VEELS derived dielectric functions.
        Theoretical curves are calculated at a reduced Lorentzian broadening of 0.1 eV.}
	\label{epself}
\end{figure}

While qualitatively correct, RPA NLF significantly overestimates the amplitude of the energy-loss peak above 40 eV, and the shoulder and peak $\sim$25 eV are slightly underestimated.
The positions of the peaks are furthermore shifted relative to experiment.
To improve oscillator strength and peak positions, LF effects (RPA LF) are applied.
The RPA LF level of theory is thus used to refine our interpretation of the m-HfO$_2$ energy-loss spectrum as well.
Fig.~\ref{epself} plots the RPA LF dielectric function together with the RPA NLF and HRTEM-VEELS derived dielectric functions.

Throughout the following discussion, we make comparisons to previously reported energy-loss spectra,\cite{Agustin,Jin,Cheynet,Couillard,Jang,ParkYang,Liou,Behar,Vos}
and in particular to earlier interpretations by Agustin et al.,\cite{Agustin} Couillard et al.,\cite{Couillard} and Park and Yang.\cite{ParkYang}

\subsection{Energy losses under 13 eV}
From the optical onset to $\sim$13 eV, which corresponds the initial slope of the energy-loss spectrum, transitions have single-particle inter-band character as evidenced by the peak in the $\Im(\varepsilon)$.
These transitions take place between the highest valence band (mostly O 2$p$ character) and the lowest conduction band (mostly Hf 5$d$ character). Experimentally, this region is highly sensitive to physical artifacts such as carbon contamination or other subbandgap defect levels, in addition to \v{C}erenkov or retardation effects\cite{Erni} and numerical artifacts due to the removal of the zero loss. Therefore this region is certainly the most difficult to access experimentally, which complicates the comparison with simulation. 

\subsection{Energy losses 13-30 eV}

At the energy resolution of 1.0--1.5 eV, the region from 13 to 16 eV appears as a shoulder.
This feature has been characterized as either inter-band transitions between O 2$p$ and Hf 5$d$\cite{Agustin} or a bulk plasmon.\cite{Couillard,Liou}
In both our TDDFT calculations and HRTEM-VEELS measurements, the $\Re(\varepsilon)$ passes through zero at $\sim$13.5 eV (dotted line in Fig.~\ref{epself}), and we thus attribute this feature to a bulk plasmon caused by the collective excitation of O 2$p$ and Hf 5$d$ and 6$s$ electrons.
As will be shown, this is the only true plasmon excitation for m-HfO$_2$.
Our theoretical predictions and experimental measurements are in remarkably good agreement for the 0-crossing that defines the plasmon energy, although the sign reversal of the real permittivity is much more pronounced in STEM rather than HRTEM experiments.

The detected oscillator strength associated with this bulk plasmon is sensitive to experimental conditions, and the resulting feature ranges from a shoulder with a similar onset energy, to a distinct peak at energies of 15 to 16 eV.\cite{Agustin,Jin,Cheynet,Couillard,Jang,ParkYang,Liou,Behar,Vos}
Some of the variation can attributed to the energy resolution of the measurement,
as we see a small peak in higher-resolution theoretical spectra that becomes a shoulder after the 1.5 eV Gaussian broadening.
We also demonstrated in previous work that the direction of momentum transfer affects peak amplitude:
the shoulder of the averaged spectrum (obtained with the same energy resolution as in this work) becomes a well-defined peak in certain directions.\cite{APL2014} 

In the range of 16 to 30 eV, the energy loss spectrum is attributed primarily to excitations of Hf 4$f$ and O 2$s$ electrons.
The Hf 4$f$ electrons are 5 eV less bound than the O 2$s$ electrons in DFT-LDA, but their combined contribution to both the $\Im(\varepsilon)$ and the energy-loss spectrum appears like a continuum.
In fact, by modeling the semicore states separately (Sec.~\ref{subsec:semicore}), we note that the net effect of Hf 4$f$ electrons is to damp energy-loss amplitude throughout this energy range.
The peak at $\sim$28 eV was previously interpreted as a plasmon.\cite{Agustin}
However, the nonzero $\Re(\varepsilon)$ indicates that this collective excitation is not a true plasmon, as is also seen in the ZrO$_2$ spectrum.\cite{Dash,Couillard}
The excitations between 16-28 eV have been previously described as inter-band transitions.\cite{Couillard,ParkYang}
However, our analysis shows that the $\Im(\varepsilon)$ decreases smoothly throughout that energy range, up to and including $\sim$28 eV where the peak is observed, while the $\Re(\varepsilon)$ remains nearly flat.
The lack of sharp features and zero-crossings in the dielectric functions leads us to conclude that between 16-28 eV,
excitations gradually transition from more single-particle character to more collective character.

\subsection{Energy losses above 30 eV}

The broad peak from 33 to 40 eV is attributed to single-particle excitations from the Hf 5$p$ electrons,
and the corresponding peak is clearly visible in the RPA NLF $\Im(\varepsilon)$.
In some experiments, this structure appears as a double peak,\cite{Agustin}
and indeed our theoretical spectrum produces a double-peaked structure as well (at $\sim$33 and $\sim$37 eV).
However we believe that this should be interpreted as a single feature.
The double-peaked structure is likely due to transition oscillator strength variations associated to the varying angular momentum character along the conduction band.
In particular, the second peak seems associated to the onset of the hybridization with Hf 6$s$ electrons on the conduction band (see Fig.~\ref{dos}).

Finally, above 41 eV, the energy-loss spectrum has a peak whose TDDFT-predicted amplitude is significantly dependent on LF effects.
In the RPA NLF energy-loss spectrum, this is the most intense peak and the $\Re(\varepsilon)$ indicates that it is the total main plasmon in m-HfO$_2$, \textit{i.e.} all electrons participate, including semicore.
However, in RPA LF, the $\Re(\varepsilon)$ is no longer negative in this region and the crossing through zero is lost.
The LF effects change this peak to a non-plasmon collective excitation that is less intense than the peak at $\sim$28 eV.
Because of the significant changes in peak amplitude in this ``LF-damped plasmon", we are unable to definitively interpret the finer features of the peak.
However, the shoulder before the peak maximum (feature G in Agustin et al.)\cite{Agustin} may be due to LF modulation of strength, since it is also visible in our RPA LF energy-loss spectrum.

\section{First-principles analysis}
\label{sec:details}

\subsection{Semicore electrons}
\label{subsec:semicore}

\begin{figure}
	\includegraphics[width=\columnwidth]{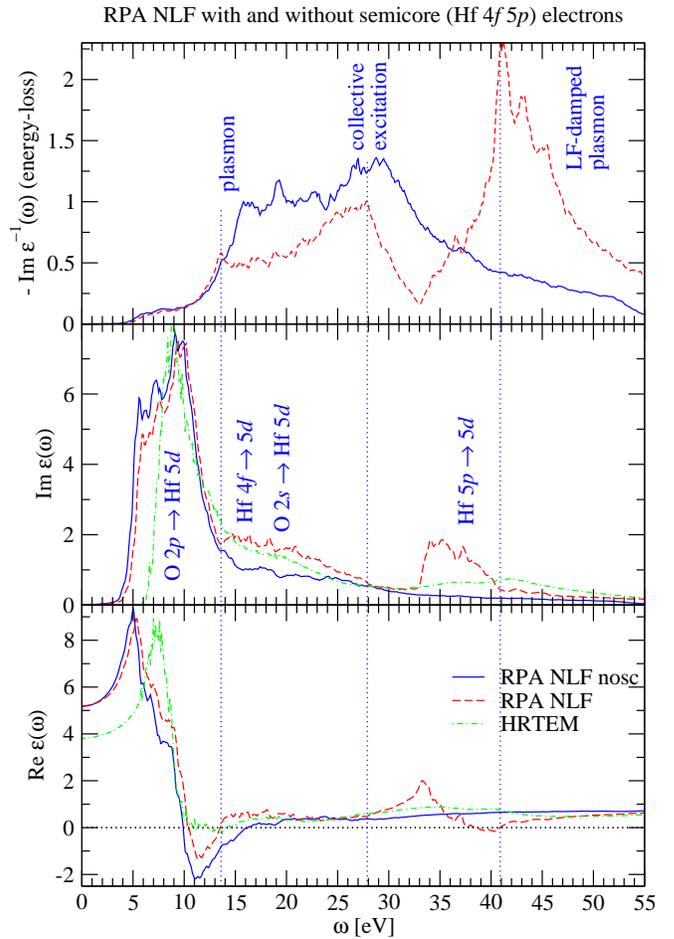}
	\caption{Energy-loss function (top), imaginary part of the dielectric function (middle), and the real part of the dielectric function (bottom) for m-HfO$_2$. 
		Solid blue line: RPA NLF without semicore (Hf 4$f$ 5$p$) electrons;
		dashed red line: RPA NLF explicitly inlcuding semicore electrons;
		and dot-dashed green line: HRTEM-VEELS derived dielectric functions.
        Theoretical curves are calculated at a reduced Lorentzian broadening of 0.1 eV.}
	\label{epselfnc}
\end{figure}

The effect of Hf semicore electrons is illustrated in Fig.~\ref{epselfnc}, where we show RPA NLF predictions of the energy-loss function, the $\Im(\varepsilon)$, and the $\Re(\varepsilon)$,
computed using two different pseudopotentials for Hf.
One pseudopotential freezes semicore electrons into the core to produce the ``RPA NLF nosc" results, while the other considers them as valence (``RPA NLF", same as in Fig.~\ref{epself}).
This comparison illustrates that the presence of semicore electrons can either increase or damp the amplitude of peaks in the energy-loss spectra.

The semicore states contribute little to the low-energy transitions ($\omega < 13$ eV), so calculations using the two pseudopotentials produce similar results in that region.
There is no noticeable difference in the energy-loss function,
although the pseudization of semicore states results in small changes in the dielectric functions.
At intermediate energies ($13\leq\omega\leq30$ eV), transitions from the Hf 4$f$ band increase the oscillator strength of the $\Im(\varepsilon)$.
As a result, the x-axis crossing of the $\Re(\varepsilon)$ shifts from $\sim$16 eV to $\sim$13.5 eV.
These relatively small changes are amplified in the resulting energy-loss function: the red-shifted plasmon peak and the entire loss function $\sim$13-30 eV is significantly damped by the additional Hf 4$f$ contributions.
At high energies ($\omega > 30$ eV), the Hf 5$p$ transitions begin.
In contrast to the featureless $\varepsilon$ obtained when semicore electrons are frozen in the pseudopotential core, the presence of the Hf 5$p$ band produces the characteristic higher-energy peak observed in m-HfO$_2$ energy-loss spectra.

\subsection{Exchange-correlation effects}
\label{subsec:manybody}

\begin{figure}
	\includegraphics[width=\columnwidth]{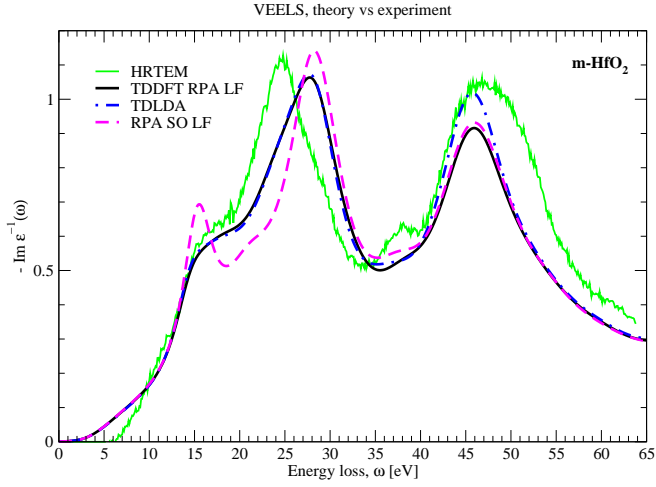}
	\caption{VEELS spectra for m-HfO$_2$. 
		Thin solid green line: HRTEM measured VEELS on a polycrystalline sample;
        solid black line: TDDFT RPA with LF effects;
		dot-dashed blue line: TDLDA;
		and dashed magenta line: RPA with the SO-corrected electronic structure.
        Theoretical curves have been convoluted with a Gaussian broadening of 1.5 eV.
		}
	\label{elfso}
\end{figure}

\begin{figure}
	\includegraphics[width=\columnwidth]{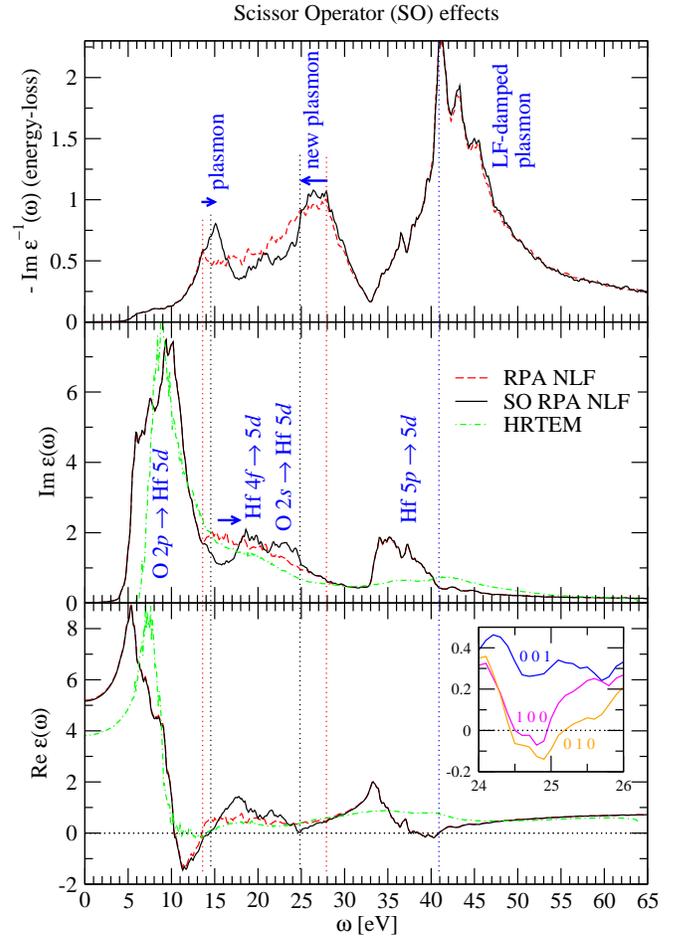}
	\caption{Energy-loss function (top), imaginary part of the dielectric function (middle), and the real part of the dielectric function (bottom) for m-HfO$_2$. 
		Dashed red line: RPA without LF effects;
		solid black line: RPA without LF effects on top of a SO-corrected electronic structure;
		and dot-dashed green line: HRTEM-VEELS derived dielectric functions.
		In the inset: RPA SO real part of the dielectric function along the main reciprocal lattice directions.
        Theoretical curves are calculated at a reduced Lorentzian broadening of 0.1 eV.
	}
	\label{epselfso}
\end{figure}

We assess the weight of exchange-correlation effects first by using TDLDA.
Typically, TDLDA energy-loss spectra exhibit only small improvements compared to RPA, and in particular at the highest energies and transferred momenta.\cite{Olevano}
That is also the case here:
the shoulder and lower-energy peak appear identical, and the higher-energy peak exhibits a slight increase of the oscillator strength that brings the amplitude into almost quantitative agreement with the experiment (Fig.~\ref{epselfso}).

We next test the explicit inclusion of electron-electron self-energy effects.
The DFT-LDA DOS (Fig.~\ref{dos}) is known to underestimate binding energies and band gaps relative to $GW$ calculations and experiment,\cite{Jiang,Bersch}
and we attempt to correct these self-energy errors by applying a SO to the Kohn-Sham DFT-LDA electronic structure.
The chosen SO decreases O 2$s$ band energies by $1.8$ eV and Hf 4$f$ band energies by $3.5$ eV to mimic the quasiparticle band structure determined from the $GW$ calculation by Jiang et al.\cite{Jiang}
This SO-corrected electronic structure is then used in a TDDFT calculation.
The resulting SO RPA energy-loss function, which includes LF effects, is shown in Fig.~\ref{elfso}, and the loss function, the $\Im(\varepsilon)$, and the $\Re(\varepsilon)$ for SO RPA NLF are shown in Fig.~\ref{epselfso}.

The most significant change after applying the self-energy SO is the increased plasmon intensity at $\sim$15 eV, which changes the shoulder in the energy-loss spectrum into a well-defined peak.
This is attributed to the blue-shifting of the Hf 4$f$ absorption peak between 15-20 eV in the $\Im(\varepsilon)$.
Because the 4$f$ transitions damp the energy-loss amplitude (Sec.~\ref{subsec:semicore}), the SO shift of Hf 4$f$ transitions to higher energies results in a plasmon excitation that is no longer damped and also blue shifted from $\sim$13.5 to $\sim$15 eV (see Fig.~\ref{epselfso}).

Initially, it may appear that this enhanced peak agrees well with the prominent plasmon peak observed in some VEELS measurements.
However, when we compare theoretical spectra to VEELS at the same energy resolution, we find that SO RPA represents the plasmon less accurately than TDLDA and RPA LF.
An analysis of SO RPA energy-loss spectra with varied momentum transfer directions shows that the intensity of this new plasmon peak is modulated, but never fully damped to a shoulder in any direction.
This contrasts with TDLDA, RPA LF, and individual HRTEM-VEELS measurements at the same energy resolution, where the plasmon appears as a well defined peak or just a shoulder depending on crystal orientation (direction of momentum transfer).\cite{APL2014}

The self-energy SO also affects the character of the peak at 25$\sim$28 eV.
The originally nearly flat averaged $\Re(\varepsilon)$ undergoes a zero-crossing at 25 eV after applying the SO.
The collective excitation is thus promoted to a real plasmon.
Again looking into contributions from various directions of momentum transfer (inset of Fig.~\ref{epselfso}), we see that the $\Re(\varepsilon)$ intersects zero for [100] and [010] momentum transfers, but not [001].
This predicted anisotropic plasmon resonance again contrasts with TDLDA, RPA LF, and HRTEM-VEELS spectra showing that a strong anisotropy is only observed on the first plasmon at $\sim$13.5 eV.\cite{APL2014}

We therefore see that the application of self-energy effects, at least as modeled using the SO, is unable to improve predictions of energy-loss spectra.
The self-energy SO-corrected electronic structure combined with RPA even produces worse predictions than RPA or TDLDA using the DFT electronic structure.
This emphasizes the need to have a balanced treatment of self-energy and excitonic interactions.\cite{Olevano,Sottile,DelSole}
Due to the cancellation of errors between these two effects in TDDFT,
simulated peaks corresponding to collective excitations have comparable amplitude to experiment.
On the other hand, differences remain between theory and experiment, such as the amplitude of the inter-band transition peak between $\sim$33-40 eV and slight shifts in peak energies.
ALDA exchange-correlation does not mitigate these differences, but we expect that the explicit inclusion of both self-energy and excitonic interactions would further improve agreement of theory and experiment.
We would also like to emphasize that the accurate treatment of many-body effects would be particularly important close to the optical band gap ($\omega<13$ eV).
In that energy range, the loss functions, which exhibit only a weak initial slope, are in good agreement, but experimental and theoretical dielectric functions present quantitative differences.

\subsection{Nonzero momentum transfer}
\label{subsec:momentum}

\begin{figure}
	\includegraphics[width=\columnwidth]{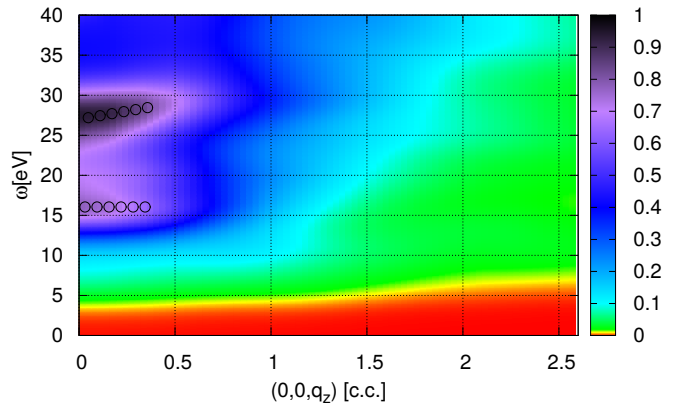}
	\caption{Energy-loss function dispersion along [001]. Open circles are a guide to the eye to indicate the dispersion of the plasmon and the collective excitation. 
    Spectra have been convoluted with a Gaussian broadening of 1.5 eV.}
	\label{dispersion}
\end{figure}

Finally, we perform TDDFT RPA LF calculations of energy-loss spectra also at non-vanishing transferred momentum to complete our picture of m-HfO$_2$ dielectric properties.
In Fig.~\ref{dispersion} we show energy-loss spectra for momentum transfers $\qv$ up to the 4th Brillouin zone ($\sim$2.5 bohr$^{-1}$) along the [001] direction.
There is a small positive dispersion of the collective excitation at~$\sim$28 eV and almost no dispersion of the plasmon at low energy.
All excitations also exhibit damping towards the Compton regime.
Similar trends are observed along the other lattice directions.
Due to the damping of excitations, anisotropies in energy-loss spectra observed at vanishing $q$ are eliminated with increasing momentum transfer.

\section{Conclusions}
\label{sec:conclusion}
We present TDDFT and HRTEM-VEELS energy-loss spectra and dielectric functions for m-HfO$_2$ and identify the excitations that result in the observed spectra.
The most prominent features of the energy-loss spectrum are the collective excitations at 13-16 eV, $\sim$28 eV, and above 40 eV.
Only the 13-16 eV feature is a plasmon.
Single-particle inter-band transitions contribute to the oscillator strength at the optical onset, from $\sim$16-28 eV, and from $\sim$33-40 eV.
By separating out the contributions of semicore electrons, we find that the Hf 4$f$ electrons damp the energy-loss oscillator strength.
Simulated spectra in the three lattice directions predict typical dispersive behavior of the collective excitations with increasing momentum transfer.

TDDFT energy-loss spectra are computed at various levels of theory, and we find that RPA and TDLDA are in good agreement with experiment as long as LF effects are included.
LF effects are found to significantly damp the peak above 40 eV, and to change the nature of the peak from a plasmon to a collective excitation.
For the same peak, the TDLDA oscillator strength is in slightly better agreement with experiment than RPA, but the two spectra do not otherwise differ.
We show that many-body effects are strongest from the optical edge through the plasmon peak, and that solely accounting for self-energy effects without compensating excitonic effects worsens agreement with experiment.
TDDFT demonstrates cancellation between these two effects through much of the energy range of interest, including for excitations from the fully-occupied Hf 4$f$ shell.
Therefore, in comparison to much more expensive many-body theory calculations, TDDFT is an efficient first-principles method to simulate and interpret VEELS.

\subsection{Acknowledgements}

We acknowledge support from the nanocharacterisation platform (PFNC) (\url{http://www.minatec.org/pfnc-plateforme-nanocaracterisation}) for experiment. 
Computer time has been provided by the French GENCI supercomputing center through project i2012096-655 and 544.

\end{document}